\documentclass[showpacs,preprint,preprintnumbers,amsmath,amssymb]{revtex4}

\usepackage{graphicx}% Include figure files
\usepackage{dcolumn}% Align table columns on decimal point
\usepackage{bm}% bold math\usepackage{natbib}
\def\upartial{\partial}

\def\R{\mbox{\textit{Ra}}}

\def\Pr{\sigma}
\def\erfc{\mathop{\rm erfc}\nolimits}

\begin{document}

%\preprint{APS/123-QED}

\title{Critical Rayleigh number of\\for error function temperature profile\\with a quasi-static assumption}% Force line breaks with \\

\author{Oliver S. Kerr}
 \email{o.s.kerr@city.ac.uk}
\affiliation{%
Department of Mathematics, City, University of London,\\
Northampton Square, London, EC1V 0HB, U.K.
}%

\date{\today}% It is always \today, today,

\begin{abstract}
When a semi-infinite body is heated from below by a sudden increase in temperature (or cooled from above) an error function temperature profile grows as the heat diffuses into the fluid. The stability of such a profile is investigated using a large-wavelength asymptotic expansion under the quasi-static, or frozen-time, approximation. The critical Rayleigh number for this layer is found to be $\R=\pi^{1/2}$ based on the length-scale $(\kappa t)^{1/2}$ where $\kappa$ is the thermal diffusivity and $t$ the time since the onset of heating.

\end{abstract}

\maketitle

\section{Introduction}

Understanding when an evolving system starts to become unstable is not always straight froward. For example, when a semi-infinite body of fluid is heated from below by impulsively increasing the bottom temperature by $\Delta T$ at an initial time $t=0$ then the heat diffuses into the fluid resulting in a temperature profile
\begin{equation}
\overline{T}=T_0+\Delta T \erfc\left(\frac{z}{2(\kappa t)^{1/2}}\right),\label{TGrad}
\end{equation}
where $T_0$ is the initial temperature, $\kappa$ the thermal diffusivity and $z$ the distance from the wall.
Intuition may be sought by looking at the region of greatest temperature gradients driving the instability, and making a comparison with results from the case of a layer between boundaries with comparable width and gradients. The stability of a horizontal layer of width $D$ with an imposed temperature difference of $\Delta T$ is determined by the Rayleigh number
\begin{equation}
\R=\frac{g\beta\Delta T D^3}{\nu\kappa}.\label{RaDef}
\end{equation}
where $g$ is the acceleration due to gravity, $\beta$ the coefficient of thermal expansion and $\nu$ the kinematic viscosity. One may think of $D$ as growing like $(\kappa t)^{1/2}$. This would lead one to expect the system to start off as stable and become unstable as the instantaneous Rayleigh number exceeded some critic value. An estimate of this value may be found by a quasi-static or frozen-time analysis where the evolving temperature profile is assumed to be fixed. This approach was used by Currie \cite{Currie67} who used a piece-wise linear approximation to the temperature profile. Using a length-scale $l$ chosen so that the added heat in the fluid is the same as that of the error function profile, the critical Rayleigh number for a semi-infinite fluid was found to be 32 for a no-slip boundary.
In comparison, for the case of a linear temperature gradient in a layer between two horizontal boundaries the critical Rayleigh numbers are $27\pi^4/4$ for stress free boundary conditions and $1707.8$ for no-slip boundaries \citep[see, for example,][]{DrazinReid}. However, the quasi-static analysis of the error-function profile given by (\ref{TGrad}) seems not have been done. The purpose of this paper is to rectify this omission.

As the instability for the semi-infinite fluid is a long-wave instability we will look at the large along-wall wavelength asymptotic limit. The solution separates into two parts. One part is near the wall on the length-scale of the thermal diffusion distance $(\kappa t)^{1/2}$. The other part represents the adjustment of the bulk of the fluid to the perturbations in this boundary layer and is on the length-scale of wavelength of the disturbances. We will match the solutions in these regions using the method of matched asymptotic expansion \citep{VanDyke,HinchEJ} and hence show the critical value of the Rayleigh number is $\R=\sqrt{\pi}$.

\section{Asymptotic analysis}

\begin{figure}
\centerline{
\includegraphics[width=1.5in]{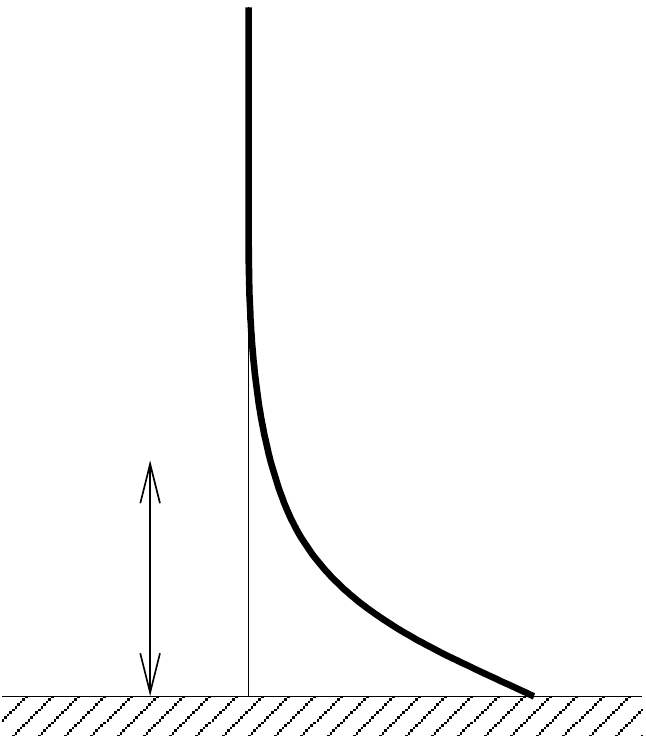}\raisebox{1.38in}{\makebox[0in][l]{\hspace{-0.8in}$T_0$}}\raisebox{0.28in}{\makebox[0in][l]{\hspace{-0.36in}$T_0+\Delta T$}}\raisebox{0.35in}{\makebox[0in][l]{\hspace{-1.65in}$(\kappa t)^{1/2}$}}\hspace{0.2in}%
} 
\caption{Schematic diagram showing the temperature profile for the heating from a single lower boundary.}\label{Schematic}
\end{figure}

Here we look at the heating of a semi-infinite body of fluid from a single horizontal boundary below the fluid. In this situation there is a growing destabilizing temperature gradient of height of order $(\kappa t)^{1/2}$ where $t$ is the time since the onset of heating and $\kappa$ the thermal diffusivity. This is shown schematically in figure~\ref{Schematic}.
If the boundary temperature is increased to $T_0+\Delta T$ and held at this level the background temperature profile, $\overline{T}(z,t)$, is given by (\ref{TGrad}).
We will restrict ourselves to looking at incompressible, two-dimensional motions, and so we can use the vorticity--streamfunction formulation.  We can derive nondimensional equations using the scalings
of $D$ for length, $D^2/\kappa$ for time, $\Delta T$ for temperature, $\kappa/D^2$ for the vorticity and $\kappa$ for the streamfunction where $D$ is a suitable length-scale.
In this quasi-static or frozen-time analysis  we use the length-scale $D=(\kappa t)^{1/2}$, and the time variation of $D$ is ignored and it is taken to be a constant. This gives the linearized nondimensional equations for the vorticity, $\omega$ streamfunction, $\psi$ and temperature, $T$,
\begin{subequations}
\begin{equation}
\frac{1}{\Pr}\frac{\upartial \omega}{\upartial t}=-\R\frac{\upartial T}{\upartial x}+\nabla^2\omega,\label{GovEqns1}
\end{equation}
\begin{equation}
\nabla^2\psi=-\omega,\label{GovEqns2}
\end{equation}
\begin{equation}
\frac{\upartial T}{\upartial t}+\frac{\upartial \psi}{\upartial x}\frac{\upartial \overline{T}}{\upartial z}=\nabla^2T.
\end{equation}
\end{subequations}
The Rayleigh number, $\R$, is the nondimensional measure of the heating, defined previously by (\ref{RaDef}), and the Prandtl number is $\sigma=\nu/\kappa$.
The nondimensional background temperature gradient is given by
\begin{equation}
\frac{\upartial \overline{T}}{\upartial z}=-\frac{e^{-z^2/4}}{\pi^{1/2}}.
\end{equation}

The boundary conditions for the disturbances that we consider are that the velocity and temperature perturbations are zero at the solid boundary at $z=0$, and tend to zero away far from the wall, giving
\begin{equation}
\frac{\upartial \psi}{\upartial x}=\frac{\upartial \psi}{\upartial z}=T=0\quad\mbox{on}\quad z=0,\quad\mbox{and}\quad \frac{\upartial \psi}{\upartial x},\,\frac{\upartial \psi}{\upartial z},\,T\to 0\quad\mbox{as}\quad z\to\infty.
\end{equation}

\begin{figure}
\centerline{%
\includegraphics[viewport=30 40 730 540,height=2.8in]{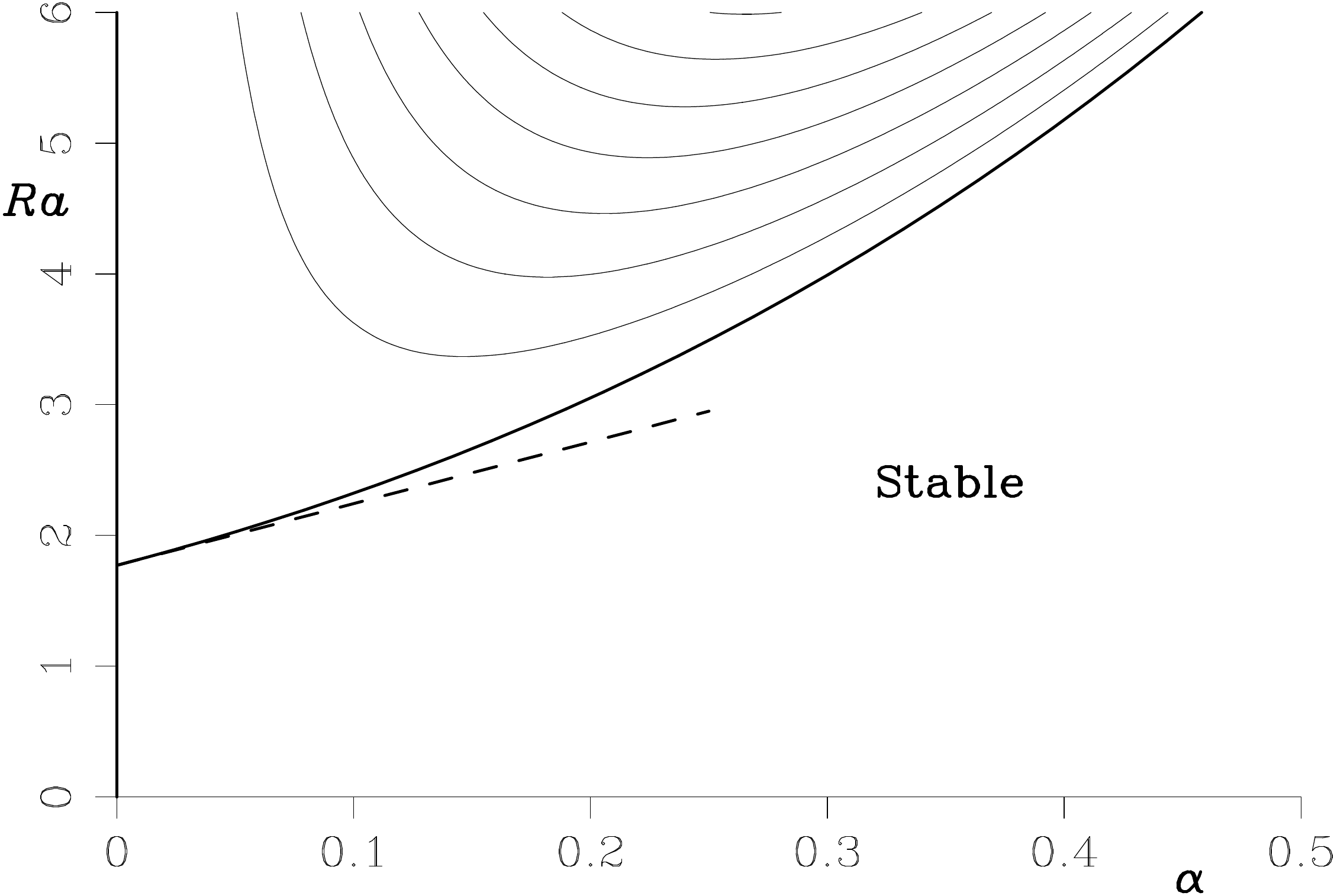}%% QSGrContPlot1.f with QS8gr_1.data
}
\caption{Stability boundary showing the critical value of $\R$ under the quasi-static approximation for the error function temperature profile. Also show are growth rate contours with intervals of $0.01$, with $\Pr=7$. The dashed line shows the first two term of the asymptotic expansion $\R=\R_0+\alpha\R_1$.}\label{QSstability}
\end{figure}

We look for solutions where
\begin{equation}
\quad \omega(x,z,t)=\omega(z)e^{\lambda t}\cos\alpha x, \ \psi(x,z,t)=\psi(z)e^{\lambda t}\cos\alpha x, \ T(x,z,t)=T(z)e^{\lambda t}\sin\alpha x,
\end{equation}
where $\omega(z)$, $\psi(z)$ and $T(z)$ are real functions.
The governing equations reduce to
\begin{subequations}
\begin{equation}
\frac{{\rm d}^2\omega}{{\rm d}z^2}=\frac{\lambda \omega}{\sigma}+\alpha^2\omega+\alpha\R  T,
\end{equation}
\begin{equation}
\frac{{\rm d}^2\psi}{{\rm d}z^2}=\alpha^2\psi-\omega.
\end{equation}
\begin{equation}
\frac{{\rm d}^2T}{{\rm d}z^2}=\lambda T+\alpha^2T-\alpha\psi\overline{T}'=\lambda T+\alpha^2T+\alpha\psi\frac{e^{-z^2/4}}{\pi^{1/2}},
\end{equation}
\end{subequations}
This system of ordinary differential equations can be solved to give the growth rate, $\lambda$, as an eigenvalue for any given $\R$, $\sigma$ and $\alpha$ using standard numerical techniques, or for marginal stability, $\lambda=0$, we can find solutions with $\R$ as the eigenvalue. Results for $\sigma=7$, a value appropriate for water, are shown in figure~\ref{QSstability}. 
The lowest curve is for marginal stability, $\lambda=0$, and is independent of the Prandtl number. The lowest point of this marginal stability curve is on the vertical axis, showing that the instability is a long-wave instability (although there is no solution to the perturbation equations for $\alpha=0$). To find the point where the curve adjoins the axis we look for an asymptotic expansion in the limit of small $\alpha$. 
We will look for the leading order behaviour in the asymptotic expansion of $\R$ for marginal stability given by
\begin{equation}
\R=\R_0+\alpha\R_1+\alpha^2\R_2+\cdots.
\end{equation}

The solution is divided into two regions. The first, the inner layer, is near the wall with $z=O(1)$, the length-scale of the temperature gradient. The second region, the outer layer, is on the scale $z=O(\alpha^{-1})$, the scale of the along-wall wavelength of the disturbances.
In the inner layer we pose the asymptotic expansions
\begin{subequations}
\begin{equation}
\omega(z)= \omega_0(z)+\alpha\omega_1(z)+\cdots,
\end{equation}
\begin{equation}
\psi(z)=\psi_0(z)+\alpha\psi_1(z)+\cdots,
\end{equation}
\begin{equation}
T(z)=\alpha \theta_0(z)+\alpha^2\theta_1(z)+\cdots,
\end{equation}
\end{subequations}
In the outer layer we use the rescaled coordinate $Z=\alpha z$, and pose the asymptotic expansions
\begin{subequations}
\begin{equation}
\omega(Z)= \Omega_0(Z)+\alpha\Omega_1(Z)+\cdots,
\end{equation}
\begin{equation}
\psi(Z)=\alpha^{-2}\Psi_0(Z)+\alpha^{-1}\Psi_1(Z)+\cdots,
\end{equation}
\begin{equation}
T(Z)=\alpha \Theta_0(Z)+\alpha^2\Theta_1(Z)+\cdots,
\end{equation}
\end{subequations}
The leading order equations in the inner layer are
\begin{subequations}
\begin{equation}
\omega_0''=0,\label{OmEqn}
\end{equation}
\begin{equation}
\psi_0''=-\omega_0,\label{PsiEqn}
\end{equation}
\begin{equation}
\theta_0''=\psi_0e^{-z^2/4}/\pi^{1/2},\label{TEqn}
\end{equation}
\end{subequations}
while the leading order equations in the outer layer are
\begin{subequations}
\begin{equation}
\Omega_0''=\Omega_0+\R_0\Theta_0,
\end{equation}
\begin{equation}
\Psi_0''=\Psi_0-\Omega_0.
\end{equation}
\begin{equation}
\Theta_0''=\Theta_0,
\end{equation}
\end{subequations}
In the outer layer the background temperature gradient term is exponentially small and so is ignored.
The boundary conditions imposed in the outer layer are that all variables decay as $Z\to\infty$. With this constraint the solutions are
\begin{subequations}
\begin{equation}
\Theta_0=A_0e^{-Z},
\end{equation}
\begin{equation}
\Omega_0=B_0e^{-Z}-\frac{\R_0 A_0}{2}Ze^{-Z},
\end{equation}
\begin{equation}
\Psi_0=C_0e^{-Z}+\frac{B_0}{2}Ze^{-Z}-\frac{\R_0 A_0}{8}\left(Z^2+Z\right)e^{-Z},
\end{equation}
\end{subequations}
where $A_0$, $B_0$ and $C_0$ are constants.

The inner layer equations (\ref{OmEqn}) and (\ref{PsiEqn}) have solutions 
\begin{subequations}
\begin{equation}
\omega_0=a_0+b_0z,
\end{equation}
\begin{equation}
\psi_0=-\frac{a_0z^2}{2}-\frac{b_0z^3}{6}+c_0+d_0z,
\end{equation}
\end{subequations}
where $a_0$, $b_0$, $c_0$ and $d_0$ are constants.
The no-slip boundary condition at $z=0$ gives $c_0=d_0=0$. We now match the streamfunction and vorticity between the layers by using an intermediate variable or Van Dyke matching rules \citep{VanDyke,HinchEJ}. This gives the requirement that the cubic term in $\psi_0$ vanishes, as must the constant and $Z$ terms in the small-$Z$ Taylor series expansion of $\Psi_0$. Hence
\begin{equation}
b_0=0,\quad C_0=0,\quad \frac{B_0}{2}-\frac{\R A_0}{8}=0.
\end{equation}
Matching the quadratic terms in the two expansions for the streamfunction gives
\begin{equation}
-\frac{a_0}{2}=-\frac{B_0}{2}\ \left({}=-\frac{\R_0 A_0}{8}\right).\label{Ratio1}
\end{equation}
Lastly, solving (\ref{TEqn}) gives
\begin{equation}
\theta_0=\frac{2a_0}{\pi^{1/2}}\left(1-e^{-z^2/4}\right)+\frac{a_0}{\pi^{1/2}}\int_0^z\int_{z''}^\infty e^{-z'^2/4}{\rm d}z'\,{\rm d}z''+e_0+f_0z,
\end{equation}
where $e_0$ and $f_0$ are constants.
As $\theta_0(0)=0$, so $e_0=0$. Matching with $\Theta_0$ in the outer layer gives $f_0=0$ and
\begin{equation}
\lim_{z\to\infty}\theta_0(z)=\frac{4a_0}{\pi^{1/2}} =A_0.\label{Ratio2}
\end{equation}
 Comparing (\ref{Ratio1}) and (\ref{Ratio2}) gives 
\begin{equation}
\R_0=\pi^{1/2},
\end{equation}
the critical value of the Rayleigh number in the limit $\alpha\to0$.
This analysis can be extended to higher orders. We find that the second term in the asymptotic expansion for $\R$ is $\R_1=3\pi/2$. The approximation given by these first two terms in this expansion is shown by the dashed line in figure~\ref{QSstability}, where we see it is a tangent to the stability boundary at $\alpha=0$, as would be expected.

\begin{figure}
\centerline{
\includegraphics[width=1.8in]{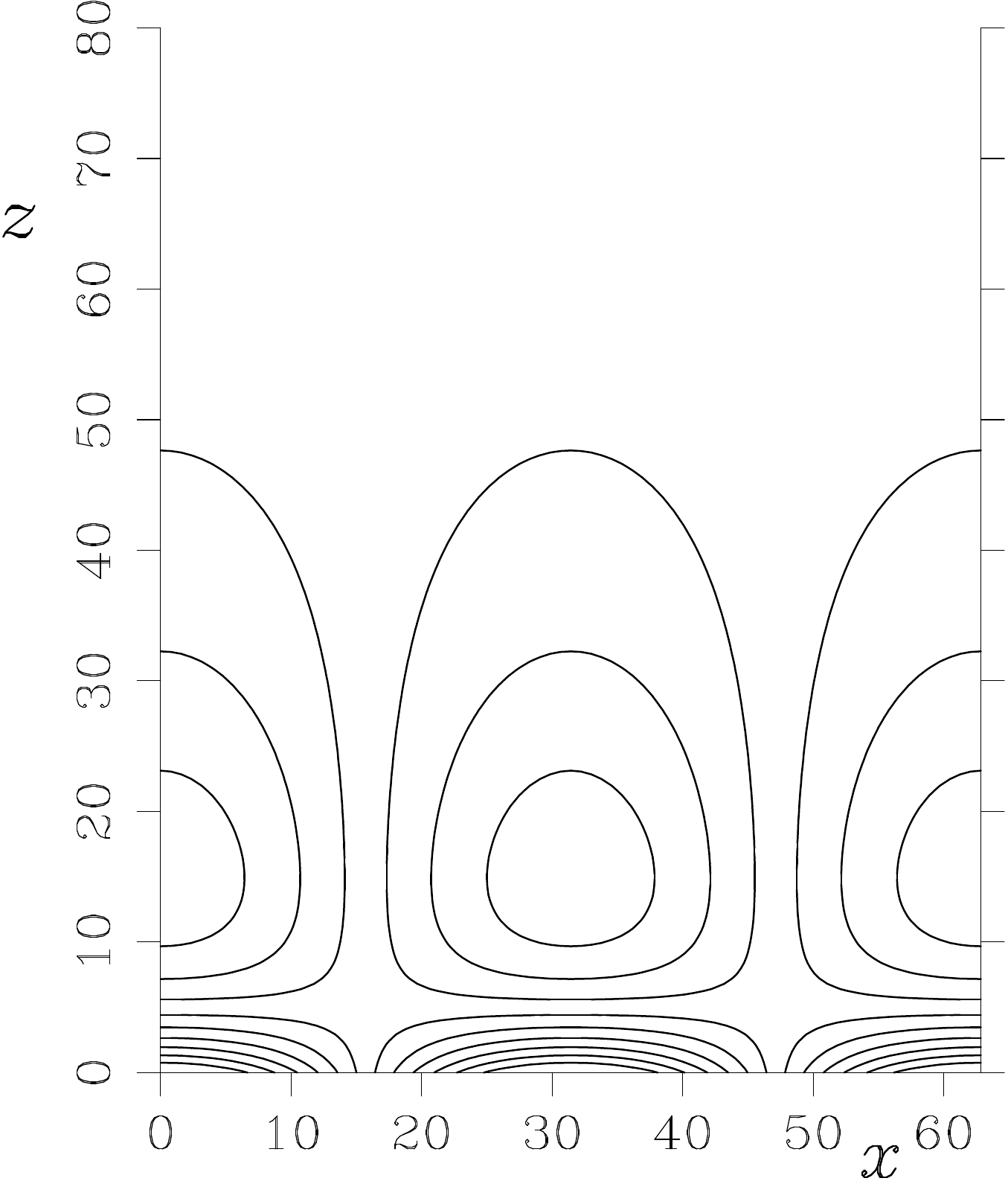}%
\hspace{0.15in}%
\includegraphics[width=1.8in]{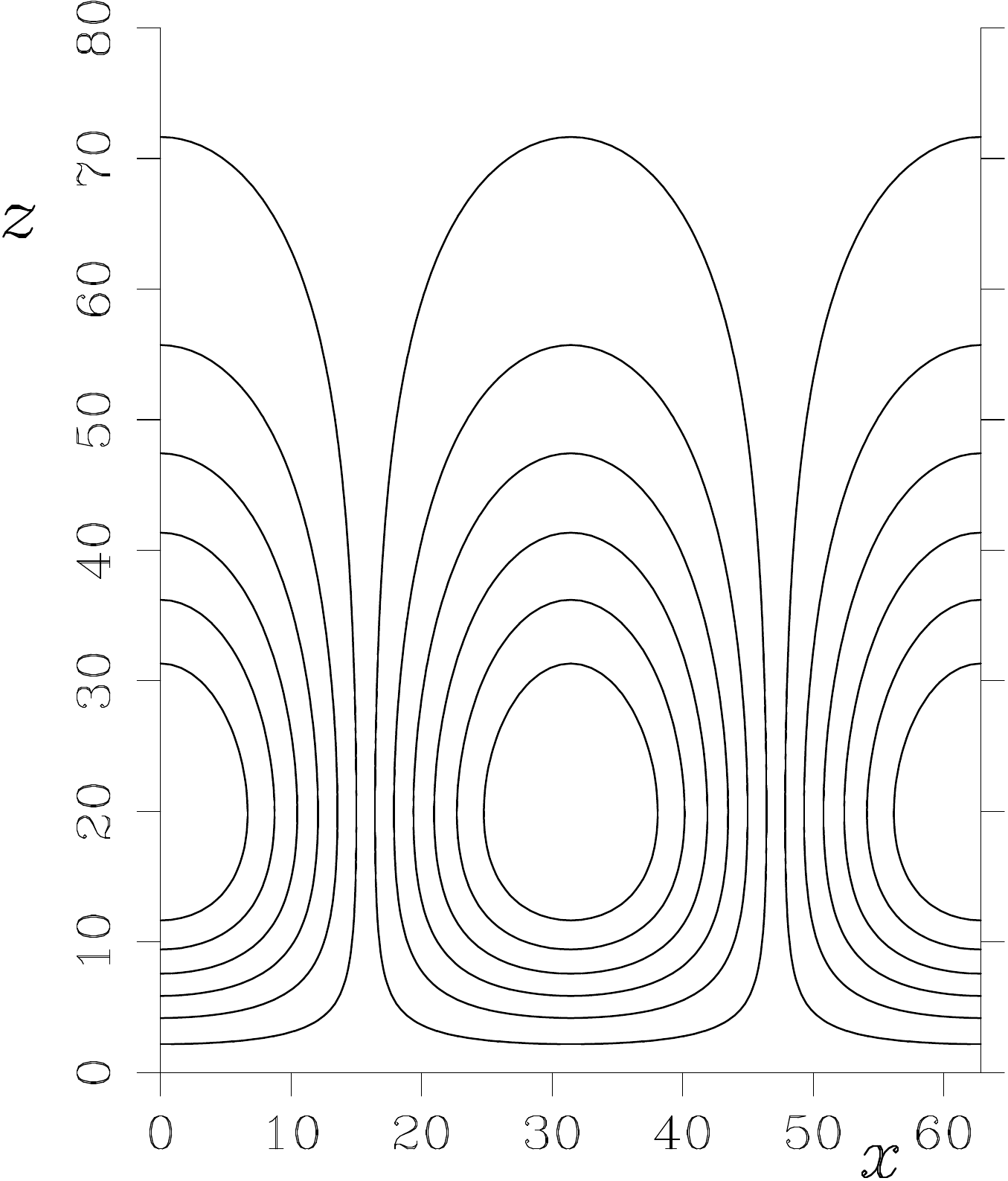}%
\hspace{0.15in}%
\includegraphics[width=1.8in]{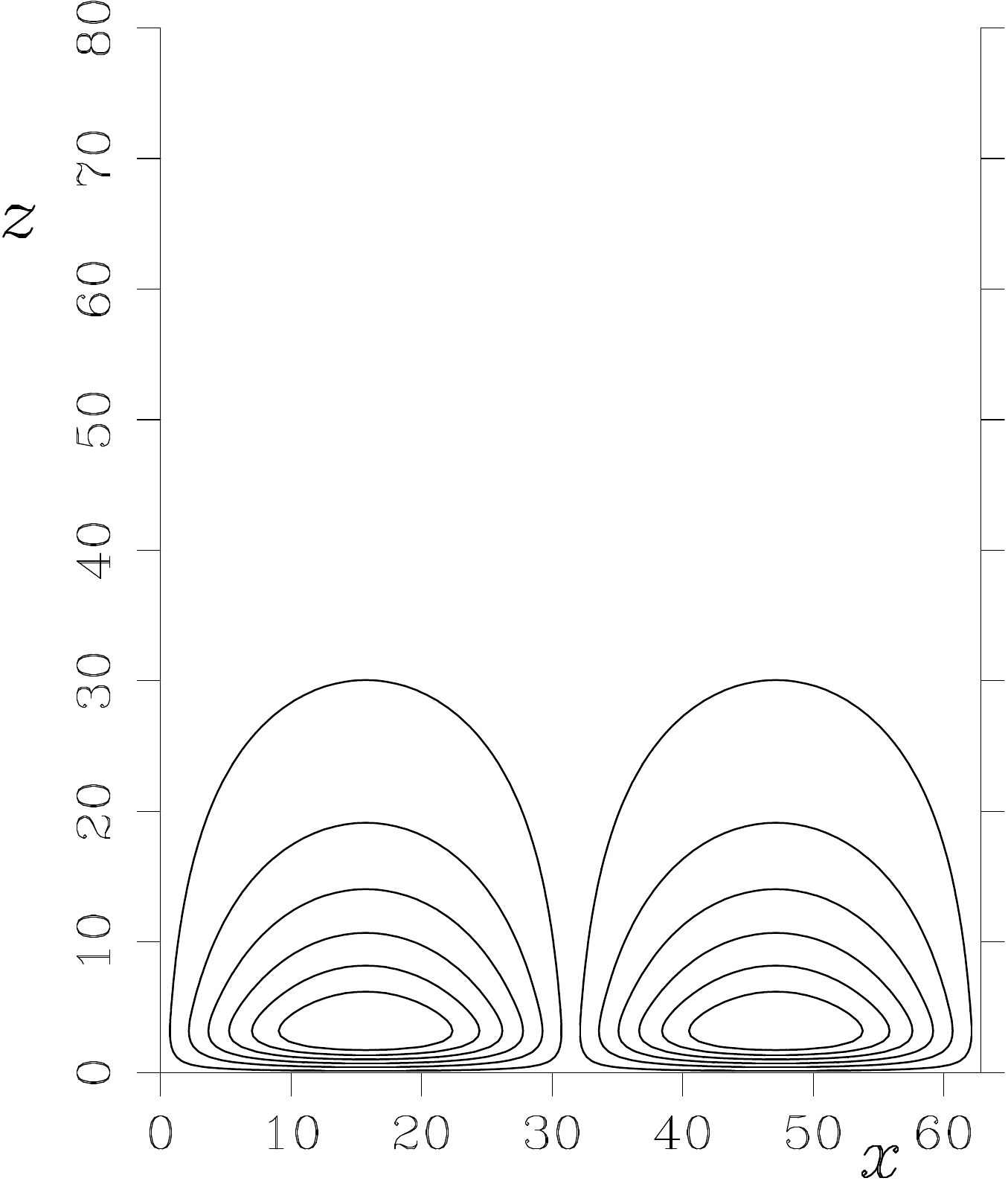}%
} 
\centerline{\hspace*{0.3in}({\it a}) \hspace{1.7in} ({\it b}) \hspace{1.7in} ({\it c})}
\caption{Contour plots of the composite asymptotic expansions for the leading order terms in the expansions for the marginally stable mode of ({\it a}) vorticity, ({\it b}) streamfunction and ({\it c}) temperature, for $\alpha=0.1$.}\label{Composite}
\end{figure}

The form of the instabilities is shown in figure~\ref{Composite}. These show the composite asymptotic expansions \citep{VanDyke,HinchEJ} of the leading order solutions of the vorticity, streamfunction and temperature which are valid in both the inner and outer layers.
The solutions to the vorticity and the streamfunctions are just the outer layer solutions which extend through the inner layer, while the temperature shows the adjustment to the zero boundary condition at the wall in the narrow inner layer.

\section{Conclusions}

We have seen here that the critical Rayleigh number for a semi-infinite body of fluid with an error function temperature profile is $\R=\sqrt{\pi}$. This analysis uses the quasi-static, or frozen-time, assumption and is based on a small-wavenumber asymptotic expansion. This resulting Rayleigh number is three orders of magnitude smaller than the critical value found for a layer with no-slip boundaries on the assumption the the boundary separation and the temperature difference across the layer are the same. The reason for this is that the instabilities in the layer are constrained to have a length-scale of the order of the layer height. This smaller scale greatly increases the effect of viscous damping.

It should be noted that, of course, the quasi-static assumption is not strictly valid for the semi-infinite body of fluid. However this same assumption is effectively being made in comparing the results of a layer whose height matches the instantaneous thickness of the thermal layer.

\providecommand{\noopsort}[1]{}

\end{document}